\documentclass{article}

\PassOptionsToPackage{numbers}{natbib}

\usepackage[final]{neurips_2024}

\usepackage[utf8]{inputenc} 
\usepackage[T1]{fontenc}    
\usepackage{hyperref}       
\usepackage{url}            
\usepackage{booktabs}       
\usepackage{amsfonts}       
\usepackage{nicefrac}       
\usepackage{microtype}      
\usepackage{xcolor}         
\usepackage{amsmath}
\usepackage{graphicx}
\usepackage{mathtools}
\usepackage{multirow}
\usepackage{multicol}
\usepackage{rotating}
\usepackage{float}
\usepackage{amsmath, amssymb}
\usepackage{caption}
\usepackage{subcaption}
\usepackage{xcolor}
\definecolor{berry}{RGB}{188,23,18}

\usepackage{enumitem}
\setlist{topsep=1pt, itemsep=1pt, partopsep=1pt, parsep=1pt}
\setlist[itemize]{align=parleft,left=0pt..1.5em}
\setlist[enumerate]{align=parleft,left=0pt..1.5em}
\usepackage[compact]{titlesec}
\titlespacing{\section}{0pt}{5pt plus 2pt minus 2pt}{3pt plus 1pt minus 1pt}
\titlespacing{\subsection}{0pt}{3pt plus 2pt minus 2pt}{2pt plus 1pt minus 1pt}
\titlespacing{\subsubsection}{0pt}{3pt plus 2pt minus 2pt}{2pt plus 1pt minus 1pt}
\title{An Experimental Study of Competitive Market Behavior Through LLMs}

\author{%
  Jingru Jia\textsuperscript{*}, Zehua Yuan\textsuperscript{*}\\
University of Illinois at Urbana-Champaign \\
  \texttt{\{jingruj3, zehuay2\}@illinois.edu}
}

\begin{document}

\maketitle
\footnotetext[1]{* Equal contribution}

\begin{abstract}
This study explores the potential of large language models (LLMs) to conduct market experiments, aiming to understand their capability to comprehend competitive market dynamics. 
We model the behavior of market agents in a controlled experimental setting, assessing their ability to converge toward competitive equilibria. The results reveal the challenges current LLMs face in replicating the dynamic decision-making processes characteristic of human trading behavior.
Unlike humans, LLMs lacked the capacity to achieve market equilibrium. 
The research demonstrates that while LLMs provide a valuable tool for scalable and reproducible market simulations, their current limitations necessitate further advancements to fully capture the complexities of market behavior. 
Future work that enhances dynamic learning capabilities and incorporates elements of behavioral economics could improve the effectiveness of LLMs in the economic domain, providing new insights into market dynamics and aiding in the refinement of economic policies.

\end{abstract}

\section{Introduction}

Competitive market behavior encompasses strategies and decision-making processes that agents employ to optimize their outcomes within market constraints, converging at an equilibrium where no participant has an incentive to deviate from their strategy. 
This state of equilibrium is critical for understanding how markets function in real-world scenarios \cite{arrow1954existence,debreu1956market,shoven1992applying}. 
Previous experiments \cite{smith1962experimental,smith1976experimental,smith1982microeconomic,smith2003constructivist} used controlled settings to replicate market conditions, allowing for observing economic principles in action. These experiments provide empirical support for market theories and help refine the predictions of economic models regarding human behavior in markets.

This study leverages LLMs's generative ability to model the dynamics of competitive markets. 
Although LLMs are trained on vast datasets that include human-like language patterns, decision-making, and even expressions of cognitive biases,\cite{jia2024decision,xue2023weaverbird,guo2023gpt,leng2024can,huang2023finbert}, it remains unclear if LLMs can exhibit the adaptive, feedback-driven behaviors observed in economic markets. This study investigates the extent to which LLMs can act as synthetic agents in a competitive market setting, offering insights into both their potential and limitations. Our focus on a double-auction experiment aims to test if LLMs can approach market equilibrium—a core concept in economic theory. Understanding this capability is essential for evaluating the applicability of LLMs in economic simulations and decision-making tasks.

We used ChatGPT-4.0 to analyze the dynamics of double auction experiments, offering an AI-driven perspective on economic market behaviors. Our findings indicate that while LLMs provide a scalable platform for simulating market dynamics, their limited adaptive learning and real-time feedback hinder their ability to reach market equilibrium, underscoring the need for further improvements to model complex economic behaviors effectively.

\section{Preliminary: Market Equilibrium and Double Auction Mechanisms}

\subsection{Conceptualizing Market Equilibrium through Welfare Maximization}

Market equilibrium represents a state where supply equals demand, and both consumer and producer surpluses are maximized. 
This is achieved through welfare optimization, where the total welfare \(W\) encompasses both consumer surplus and producer surplus.

Let \(D(p)\) and \(S(p)\) represent the demand and supply function, with \(p_e\) as the equilibrium price and \(p_{max}\) as the maximum price consumers are willing to pay, the total welfare \(W\) can be expressed as:
\[
W = \int_{p_e}^{p_{max}} D(p) \, dp - \int_0^{p_e} S(p) \, dp
\]
To find the equilibrium price \(p_e\) that maximizes \(W\), the following optimization problem is solved:
\[
\max_{p_e} \left( \int_{p_e}^{p_{max}} D(p) \, dp - \int_0^{p_e} S(p) \, dp \right)
\]

Market equilibrium is achieved when the aggregated demands of buyers and supplies of sellers are matched at a price level that clears the market. This involves adjustments in prices and quantities until no incentive exists for participants to change their prices or quantities.

\subsection{Double Auctions and Market Dynamics}

\textbf{Double Auction Mechanism:}

Let \(B = \{b_1, b_2, \ldots, b_n\}\) represent the set of bids and \(A = \{a_1, a_2, \ldots, a_m\}\) represent the set of asks submitted in the market, where \(n\) and \(m\) are the number of bids and asks. Each bid \(b_i\) and ask \(a_j\) can be considered as order pairs \((p_i, q_i)\) and \((p_j', q_j')\), where \(p\) and \(q\) denote the price and quantity.

\begin{itemize}
    \item \textbf{Transaction Rule:} A transaction occurs whenever \(b_i \geq a_j\). The transaction price \(p^*\) is typically set at \(b_i\) or \(a_j\), depending on who called the price first.
    \item \textbf{Market Clearing Condition:} The market is considered to be in equilibrium when there are no further bids higher than any asks, effectively clearing the market.
\end{itemize}

\textbf{Equilibrium Analysis:}

In a perfectly competitive market, the equilibrium price \(p_e\) aligns with the supply and demand functions. Assuming linear supply \(S(p) = sp + k\) and demand \(D(p) = tp - h\) functions, where \(s, k, t,\) and \(h\) are constants, equilibrium occurs at \(p_e\) where \(S(p_e) = D(p_e)\). That is:
\[
sp_e + k = tp_e - h \implies p_e = \frac{h - k}{s - t}
\]

\section{Experiment Design}\label{sec:exp}

The experimental setup closely follows the configuration of the original experiment series \#1 from the foundational market equilibrium study \cite{smith1962experimental}. 

\textbf{Participant Setup: } 

A total of 22 agents—11 buyers and 11 sellers—were deployed in each trading session, with each agent randomly assigned either a buyer or seller role at the beginning of the experiment. Each participant receives a card with a number between 0.75 and 3.25, indicating either the cash they hold (for buyers) or the cost of their item (for sellers). Sellers can only list prices above their cost, while buyers can only bid below their available cash.

\textbf{Market Structure and Trading Rules:}

The market operated under a double-auction framework, where both buyers and sellers submitted their respective bids and asks. The experiment adhered to the following rules and mechanisms: 

\textit{Bids and Asks:} Each buyer posted a bid (maximum price they’re willing to pay) and each seller posted an ask (minimum price they’re willing to accept) based on their assigned budgets and costs.
\textit{Matching Algorithm:} A transaction occurred when a buyer’s bid met or exceeded a seller’s ask. Each trading session consisted of five rounds, designed to emulate a trading day, with cumulative information provided to agents as the rounds progressed. 

\begin{figure}
    \centering
    \includegraphics[width=0.8\linewidth]{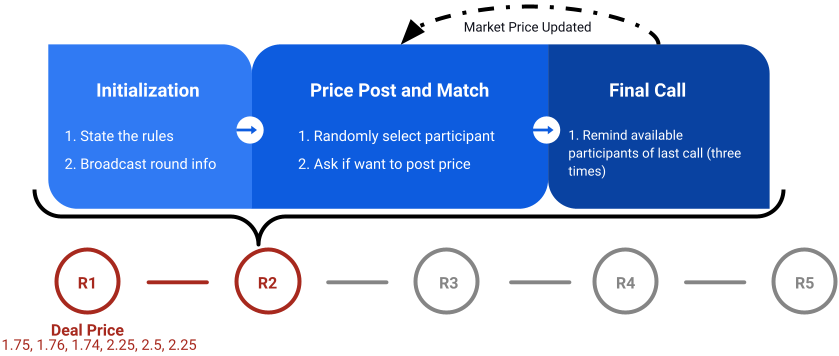}
    \caption{Baseline Experiment Three-step Process}
    \label{fig:exp_process}
    \vspace{-1em}
\end{figure}

To adapt this process for use with LLMs, specifically ChatGPT-4.0, we designed a three-step approach in Figure \ref{fig:exp_process}to abstract the procedures used in human experiments. This method ensures the experiments are conducted in a formal and reproducible manner.

\textbf{Step 1: Initialization Stage}  
At the start of each round, the system sent a structured prompt to each participant (representing a separate ChatGPT-4 session) detailing the double-auction rules and providing specific information about the current round. Participants were required to confirm their understanding of the rules before proceeding. For subsequent rounds, only the information about the previous transaction was updated in the prompt, while the remainder of the message remained unchanged, allowing agents to make decisions based on recent transaction outcomes without altering the core instructions.

\textbf{Step 2: Price Posting and Matching}  
To emulate spontaneous price-posting behavior seen in human markets, a random session was selected using a number generator. Depending on the selected session’s role, three types of prompts were issued: a. If a buyer posted a price, b. If a seller posted a price, or, c.If both a buyer and a seller posted prices. If a participant chose to post a price, their response was recorded. A transaction occurred when a seller's price was equal to or lower than a buyer's price, or vice versa. The agreed-upon transaction price was then recorded and added to the transaction history, which would be shared with participants in the subsequent round, providing them with recent market feedback.

\textbf{Step 3: Final Call}  
In cases where participants did not respond to initial system prompts to post or update their prices, a final call reminder was issued to all agents. If there were still no updates after three final call reminders, the trading day was deemed concluded. This iterative process was repeated over five rounds per session, consistent with the methodology of the original double-auction studies.

\section{Result and Discussion}

Data on all bids, asks, and executed transactions were recorded in real-time. Key metrics included: \textbf {a. Price Convergence.} Tracking how closely transaction prices approached the theoretical equilibrium price over the course of multiple rounds. \textbf{b. Volatility.} Measuring fluctuations in transaction prices and bid-ask spreads to assess market stability. \textbf{c. Agent Strategy Adaptation. }Observing changes in agent behaviors, such as adjustments in bidding and asking strategies, to assess if LLMs displayed any adaptive decision-making similar to human participants.

\textbf{Summary of Findings:}

\begin{figure}
    \centering
    \includegraphics[width=0.4\linewidth]{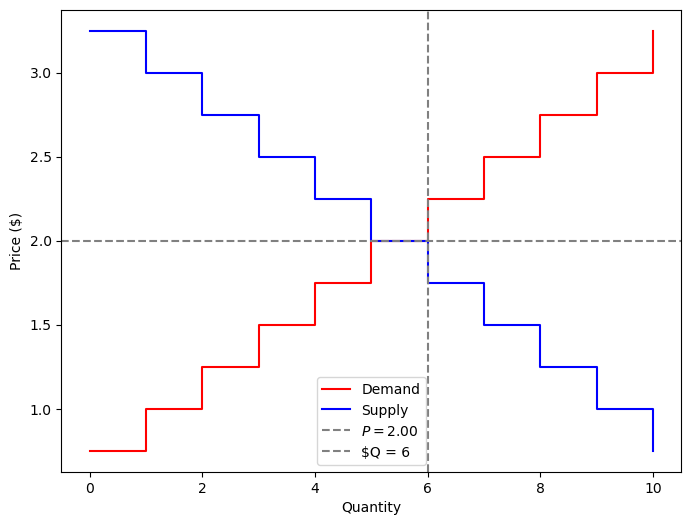}
    \caption{Demand and Supply Curves}
    \label{fig: Demand and Supply Curves}
    \vspace{-1em}
\end{figure}

\begin{figure}[h]
    \centering
    \includegraphics[width=\linewidth]{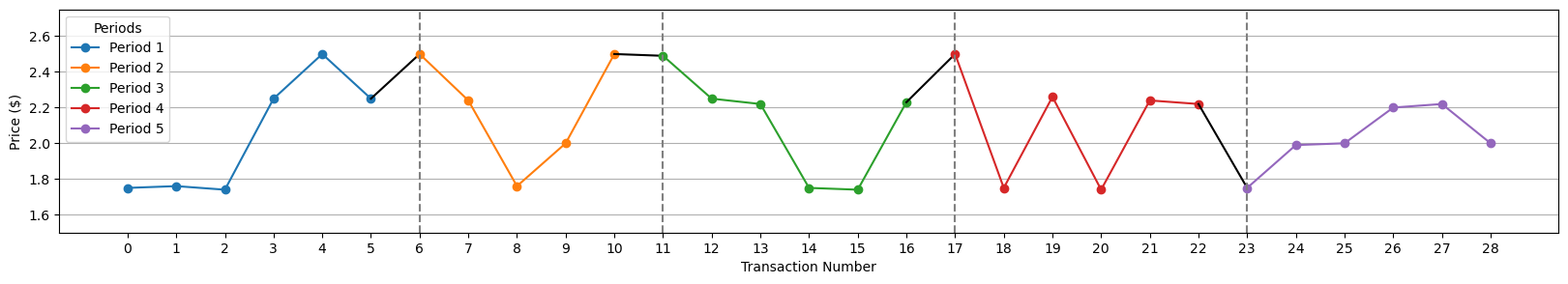}
    \caption{Transaction Price}
    \label{fig:Transaction Price}
\end{figure}

Figure \ref{fig: Demand and Supply Curves} the predetermined demand and supply curves for the LLMs in our experiments. Based on the procedures outlined in Section \ref{sec:exp}, we documented transaction prices and trends as depicted in Table \ref{tab: result} and Figure \ref{fig:Transaction Price}. In experiments involving human subjects \cite{smith1962experimental}, transaction prices typically display a gradual trend of convergence toward the theoretical equilibrium price of \$2.00. This behavior aligns with the Walrasian equilibrium theory \cite{vega1997evolution}, which posits that in a market of rational agents who are fully informed and act to maximize utility, prices will adjust to balance supply and demand, effectively clearing the market. 

\begin{table}[h]
\begin{tabular}{cccccc}
\hline
Trading Period &
  \begin{tabular}[c]{@{}c@{}}Predicted \\ exchange \\ quantity\end{tabular} &
  \begin{tabular}[c]{@{}c@{}}Actual \\ exchange \\ quantity\end{tabular} &
  \begin{tabular}[c]{@{}c@{}}Predicted \\ exchange \\ price\end{tabular} &
  \begin{tabular}[c]{@{}c@{}}Average \\ actual \\ exchange price\end{tabular} &
  \begin{tabular}[c]{@{}c@{}}Coefficient \\ of convergence\end{tabular} \\ \hline
1 & 6 & 6 & 2 & 2.04 & 4.17  \\
2 & 6 & 5 & 2 & 2.2  & 20.00 \\
3 & 6 & 6 & 2 & 2.11 & 11.33 \\
4 & 6 & 6 & 2 & 2.12 & 11.83 \\
5 & 6 & 6 & 2 & 2.02 & 2.67  \\ \hline
\end{tabular}
\caption{Result of the Transaction Prices}
\label{tab: result}
\vspace{-1em}
\end{table}

However, in the LLM-driven experiment, this pattern of convergence toward the equilibrium price was not observed. 
Instead, the transaction prices generated by the LLMs fluctuated around the target equilibrium price of \$2.00 throughout the trading periods, without a clear trend toward stabilization. 
In our experiment, the first trading period had a coefficient of convergence of 4.17, suggesting a moderate deviation from the equilibrium. However, subsequent periods, such as the second, displayed a much higher deviation with a coefficient of 20.00, reflecting substantial fluctuations in pricing decisions made by the LLM. This variability persisted throughout the experiment, with the fifth and final period showing a slight improvement in convergence (a coefficient of 2.67), though still not indicative of a stable trend towards equilibrium. 
This lack of convergence in LLMs suggests that, despite being trained on vast datasets with human-like language and behavioral patterns, LLMs operate with fundamentally different mechanisms. Human participants use iterative learning to adjust their bids and asks based on past outcomes, whereas LLMs lack the capacity to adapt across rounds. Several factors inherent to the nature of LLMs and their operational framework might explain this discrepancy:

\textbf{Lack of Adaptive Learning:} In human trading, adaptive learning can be modeled using a recursive updating of price estimates formula \cite{carceles2007adaptive, berardi2017empirical}:
\[
p_{t+1} = p_t + \gamma (p^* - p_t)
\]
where \( p_t \) is the current price at time \( t \), \( p^* \) is the transaction price observed, and \( \gamma \) is the learning rate indicating how quickly agents incorporate new information. In contrast, LLMs may use a static model: $p = f(x)$, where \( p \) is derived directly from inputs \( x \) without adjustment over time.

\textbf{Algorithmic Rigidity and Limited Market Feedback Integration:} Human traders dynamically adjust their strategies using a model \cite{berardi2017empirical, carceles2007adaptive}:
$ p_t = \alpha S_t + \beta D_t + \epsilon$, 
where \( S_t \) and \( D_t \) represent supply and demand at time \( t \), and \( \alpha \) and \( \beta \) adapt over time. They also update their strategies based on ongoing market feedback, modeled by \cite{narayan2011peer, kyburg1987bayesian}:$\pi_{t+1}(p) = \frac{\pi_t(p) \cdot L(p; y_t)}{\int \pi_t(p) \cdot L(p; y_t) dp}$, where \( \pi_t(p) \) is the prior belief about price, and \( L(p; y_t) \) is the likelihood of observing a signal \( y_t \). Conversely, LLMs may use a static approach with fixed parameters like:
$p = \theta^T x$, and a simple feedback mechanism: $p = g(y_t, \theta)$, where \( \theta \) remains constant, combining adaptation and feedback into a single, less flexible process. In the experiment, while LLMs were provided with transaction information from previous rounds, their responses did not adjust based on these historical outcomes. Thus, their behavior remained relatively invariant across rounds, resulting in volatile price movements rather than converging towards a market-clearing price.

\textbf{Absence of Psychological Factors:} Human agents bring cognitive biases, emotional influences, and bounded rationality into economic decisions, which often facilitate convergence towards equilibrium by introducing adaptive variability. For instance, human traders frequently display loss aversion and risk-seeking or risk-averse behaviors that influence their strategies. This behavior can be modeled using the following utility function: $U(x) = x^\alpha \cdot \exp(-\lambda \cdot \delta)$, where \( x \) represents the monetary outcome, \( \alpha \) captures risk aversion, \( \lambda \) is the loss aversion coefficient, and \( \delta \) is the deviation from a reference point. LLMs lack such complex emotional frameworks, operating instead within the confines of utility maximization devoid of emotional biases: $U(x) = \sum w_i x_i$, where \( w_i \) are weights assigned to different outcomes \( x_i \), calculated without behavioral adjustments.

\section{Conclusion}

This study has demonstrated significant differences between the behaviors of LLMs and human traders in achieving market equilibrium, building on the foundational experimental economics work \cite{smith1976experimental}. While LLMs offer consistency and lack the biases often seen in human behavior, they also lack the dynamic adaptability and feedback-driven decision-making that human traders possess. This limitation hinders their ability to achieve Walrasian equilibrium and underscores the need for more sophisticated models capable of adjusting to changing market conditions in real time. By systematically evaluating LLMs in a competitive market environment, this study highlights pathways toward enhancing their effectiveness through targeted technological and methodological advancements.

\textbf{Key Contributions:}
This research contributes to the growing field of LLM-driven economic simulations by providing one of the first evaluations of LLMs’ capabilities in modeling complex market behaviors. It establishes a baseline understanding of how LLMs function within economic contexts, and offers insights into their current limitations and opportunities for improvement. This work lays the groundwork for future studies to refine LLMs, equipping them with adaptive learning mechanisms and behavioral economics elements to better approximate human decision-making processes in markets. By focusing on the double-auction setting, a well-established economic experiment, this study also bridges the gap between traditional experimental economics and the emerging applications of LLMs in social science.

The insights gained from this study have significant implications for both research and business applications. In research, LLMs equipped with adaptive learning could be utilized to model complex economic scenarios, such as the effects of policy interventions or changes in consumer behavior, without the need for large-scale human-based studies. This capability would be valuable for fields like experimental economics, behavioral finance, and macroeconomic forecasting. In business, enhanced LLMs could serve as virtual agents to model consumer behavior in dynamic pricing, auction platforms, or market forecasting. Such applications would enable companies to test pricing strategies, product placements, and other decisions in a simulated environment, reducing risk and optimizing outcomes before applying changes in real markets.

\textbf{Future Directions and Implications:} 

Incorporating adaptive learning and behavioral economics could enhance the utility of LLMs as economic-related agents or economic research. Such models could also facilitate large-scale simulations to test economic theories and predict market responses to regulatory changes with high precision. This research serves as an initial evaluation of LLMs' capabilities in modeling market behavior. Future work should explore diverse settings, such as varying market conditions, different types of auction mechanisms, and enhanced LLM training protocols, to fully assess the potential and limitations of LLMs in economic simulations.


\newpage
\bibliographystyle{splncs04}
\bibliography{main}







\newpage

\newpage
\appendix

\end{document}